\documentclass[aps,prl,reprint,groupedaddress]{revtex4-1}
\usepackage{graphicx}
\usepackage{braket}
\usepackage{amsbsy}
\usepackage{amsmath}
\usepackage{bm}
\usepackage{euscript}
\begin{document}

\title{Qualitative Insight and Quantitative Analysis of the Effect of Temperature on the Coercivity of a Magnetic System}

\author{Mariia Moskalenko$^{1,2}$, Pavel F. Bessarab$^{3,4}$, Valery M. Uzdin$^{2,4}$, and  Hannes J\'onsson$^{1,5}$}
\affiliation{$^1$Science Institute and Faculty of Physical Sciences, Univ. of Iceland, 107 Reykjav\'{\i}k, Iceland}
\affiliation{$^2$St. Petersburg National Research University of Information Technologies, Mechanics and Optics, St. Petersburg, 197101, Russia }
\affiliation{$^3$Dept. of Materials and Nanophysics, Electrum 229, Royal Institute of Technology (KTH), SE-16440 Kista, Sweden}
\affiliation{$^4$Department of Physics, St. Petersburg State University, St. Petersburg, 198504, Russia}
\affiliation{$^5$Department of Applied Physics, Aalto University, Espoo, FIN-00076, Finland}

\date{\today}
\begin{abstract}

The temperature dependence of the response of a magnetic system to an applied field can be understood qualitatively by considering variations in the energy surface characterizing the system and estimated quantitatively with rate theory.
In the system analysed here, Fe/Sm-Co spring magnet, the width of the hysteresis loop is reduced to a half when temperature is raised from 25~K to 300~K. This narrowing can be explained and reproduced quantitatively 
without invoking temperature  dependence of model parameters as has typically been done in previous data analysis. The applied magnetic field lowers the energy barrier for reorientation of the magnetization but thermal activation brings the system over the barrier. A 2-dimensional representation of the energy surface is developed and used to gain insight into the transition mechanism and to demonstrate how the applied field alters the transition path. Our results show the importance of explicitly including the effect of thermal activation when interpreting experiments involving the manipulation of magnetic systems at finite temperature.

\end{abstract}


\maketitle



The stability of magnetic states with respect to thermal fluctuations and external perturbations is an important problem in 
fundamental studies of magnetism and of critical importance in the design of nanoscale magnetic devices~\cite{Braun12}.
A theoretical estimate of the thermal stability of magnetization in nanoscale structures as a function of size and shape 
could help design devices with unprecedented capacity and functionality.   
The preparation of a magnetic system in a particular state can be destroyed by thermally activated transitions to other available states, contributing to the temperature dependence of various properties of magnetic materials, including hysteresis loops~\cite{sharrock_94,goll_07,Suess07,Suess11,maaz_10,fernandez_05,paulus_01,sellmyer_01,ruigrok_00}. 
If the energy barrier separating stable states is large
compared to thermal energy, the thermally activated transition become a rare event and special techniques are required for the long time scale spin dynamics simulations~\cite{xue_00,chubykalo_05,chubykalo_00,Suess11}. 
According to statistical rate theories, the rate of a thermally activated, over-the-barrier transition typically follows Arrhenius law~\cite{Brown79,Coffey01}, which contains two parameters: 
an
energy barrier for 
the transition and a pre-exponential factor. While 
several calculations of an energy barrier for a magnetic transition have been reported 
~\cite{dittrich_02,dittrich_03,thiaville_03,e_03,dittrich_04,dittrich_05,suess_06,Suess07,berkov_07, goll_07,krone_10,visscher_12,fiedler_12,bessarab_12,tudosa_12,bessarab_13,bessarab_14}, 
only a few calculations of a pre-exponential factor for magnetic transitions involving
many degrees of freedom have been presented~\cite{bessarab_12,bessarab_13,fiedler_12}. 
In this  
article,
we analyze the effect of thermal activation on hysteresis loops  
using HTST
where  
both the energy barrier and the pre-exponential factor are 
evaluated to obtain an estimate of the time scale of the transition.

Here, we focus on spring magnets which are formed by combining a soft and hard magnetic material.
The combination of the large magnetization of the soft magnetic layer and the high coercitivity of the hard magnetic layer
makes these systems suitable for building permanent magnets with a large energy product (magnetic saturation times coercivity)~\cite{Goto65,Kneller91,Skomski93,Mibu96,Wuchner97}. 
Such systems can have various practical applications,
for example, in temperature assisted magnetic recording~\cite{Thiele03}.
Furthermore, spring magnets have become important test systems for the study of non-collinear magnetism and the way
magnetic field can affect the ordering of magnetic moments at the atomic scale.

\begin{figure}[h!]
\centering
\includegraphics[width=\columnwidth]{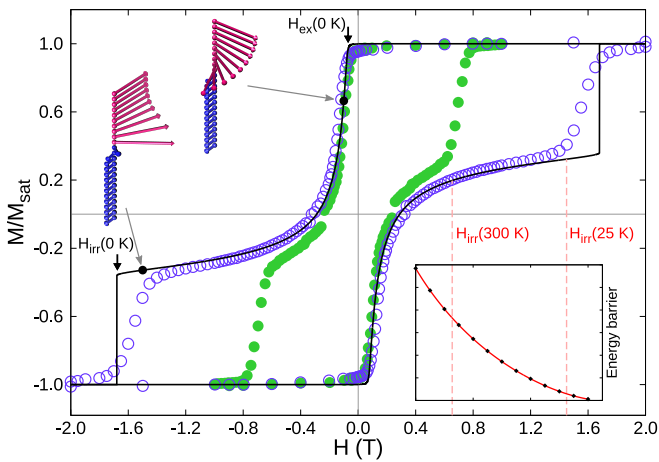}
\caption{
Magnetization of a Fe/Sm-Co spring magnet relative to saturation. Experimental results for 
25 K and 300 K (shown with open and filled circles, resp.) are taken from ref.~\cite{Fullerton99}. 
Calculated results (solid line) are obtained by minimizing the energy of the model given by Eq.~(\ref{equ:model}) and
show the predicted hysteresis loop at zero temperature.
The lower right inset shows calculated activation energy as a function of applied magnetic field obtained from the minimum 
energy path between the non-collinear spiral state and the homogeneous state with parallel moments. 
The dashed red lines show the calculated values of H$_{irr}$ at 25 K and 300 K which are in close agreement with
the experimental observations.
The insets in the upper left part show the orientation of the magnetic vectors at two points along the hystersis curve (marked 
with filled circles).
}
\label{fig1}
\end{figure}

Fig.~\ref{fig1} shows experimental data reported for a thin film spring magnet consisting of a 
20 nm layer of Sm-Co hard magnet and 20 nm Fe soft magnet~\cite{Fullerton98,Fullerton99}. 
The width of the hysteresis loop is reduced to less than a half by increasing the 
temperature from 25~K to 300~K.  
The reduction of the measured width of the hysteresis loop was ascribed to a reduction in the anisotropy parameter of the model
with temperature~\cite{Fullerton98,Fullerton99}. 
Similar data and interpretation have been reported for various other systems, see for example~\cite{Gu10}. 

In the analysis presented here, 
a 2-dimensional representation of the multidimensional energy surface is developed to
illustrate the effect of the applied field and to explain how temperature affects the hysteresis loop.
We, furthermore, show that harmonic transition state theory (HTST) 
for magnetic transitions~\cite{bessarab_12,Bessarab13a} combined with an efficient way to 
find the minimum energy path (MEP) of a transition~\cite{NEB,Bessarab15} 
can in a rather straightforward way be used to reproduce quantitatively the narrowing of the hysteresis loop 
using the same multilayer model as in refs. \cite{Fullerton98,Fullerton99} but without introducing a
temperature dependence of the model parameters. While MEPs are independent of temperature, transition rates estimated from MEPs and HTST are temperature dependent 
as in the
Arrhenius law.

When a magnetic field is applied to a spring magnet and increased gradually, 
a non-collinear spiral structure first develops in the soft magnetic material. 
The film is ideally thin enough to avoid grain boundaries in the normal direction 
and the system can return to the initial state if the field is removed.
At a large enough
field strength, however,
the magnetization of the hard magnetic material is irreversibly switched, as illustrated in Fig.~\ref{fig1}. 

The hysteresis loop of a bilayer spring magnet can contain two characteristic jumps in the magnetization~\cite{Fullerton98,leineweber_97}, as seen in Fig.~\ref{fig1}.
At a magnetic field denoted as H$_{ex}$, the spiral magnetic structures starts 
to form in the soft magnetic layer (see insets in Fig.~\ref{fig1}). As the field strength increases, the magnetic moments in the
spiral rotate further towards the direction of the applied field. 
Eventually, at a field of H$_{irr}$, the magnetization of the hard magnetic layer is reversed.  
Between H$_{ex}$ and H$_{irr}$, the stretching of the spiral structure is reversible. 

Measured hysteresis loops of spring magnets have been reproduced successfully using a multilayer model where 
each atomic layer in the film is represented with a magnetic moment vector~\cite{Fullerton98}.
The energy per unit area of the system is given by a sum of terms representing exchange interaction between adjacent layers, easy-axis anisotropy in 
the
plane of the atomic layers and interaction with an external field~\cite{visscher_07,dobin_06}
$$
E \  = \ -  \sum_{i=1}^{N-1} {\frac{A_\alpha} {d}} \cos{(\phi_i - \phi_{i+1})} \ - \ 
d \sum_{i=1}^{N} K_\beta \cos^2{(\phi_i)} \ - \\
$$
\vskip -0.7 true cm
\begin{eqnarray}
\label{equ:model}
\ \ \   \ \ \ \ \ \ \ \ \ \ \ \ \ \ \  - \  d \sum_{i=1}^{N} H M_\beta  \cos{(\phi_i - \phi_H)} 
\end{eqnarray}
Here, $\phi_i$ and $\phi_H$ the angles the magnetic 
moment vector in layer $i$ and the external field, respectively, make with the anisotropy axis. In order to determine the state of the 
system at a given field, the energy is minimized with respect to 
the 
$\phi_i$ for all layers.  
In Eq.~(\ref{equ:model}), $M_\beta$ and $K_\beta$ are the magnetization and anisotropy constant, respectively, in the soft ($\beta=s$) and hard ($\beta=h$) magnet, 
$d$ is the distance between atomic layers, and $A_\alpha$ the exchange coupling between adjacent layers in the soft magnet ($\alpha=s$), hard magnet ($\alpha=h$) and at the interface ($\alpha=int$).  
The values of these parameters for the hard magnet and the soft magnet as well as the coupling between the two
were chosen to be the same as those used in previously reported fits to the experimental data taken at 25 K~\cite{Fullerton98}
except for a 20\% increase in $K_h$ to make the model consistent with zero temperature.
Even at 25 K, thermal activation needs to be taken into account, and the fitted parameter values 
in ~\cite{Fullerton98}
then implicitly contain the effect of 
temperature on H$_{irr}$. 
We, however, explicitly include this temperature effect and need model parameters that   
correspond to
zero temperature.

The model described by Eq.~(\ref{equ:model}) defines an energy surface for the system as a function of the orientation
of the magnetic moment vectors, the $\phi_i$ variables. 
The non-collinear spiral state is a local minimum on this energy surface 
which can be found by minimizing the energy with respect to the $\phi_i$ variables.  When a significant field is applied, 
another
deeper
minimum corresponds to the collinear state where all the magnetic vectors are pointing in the direction of the field.

\begin{figure}[h!]
\centering
\includegraphics[width=0.9\columnwidth]{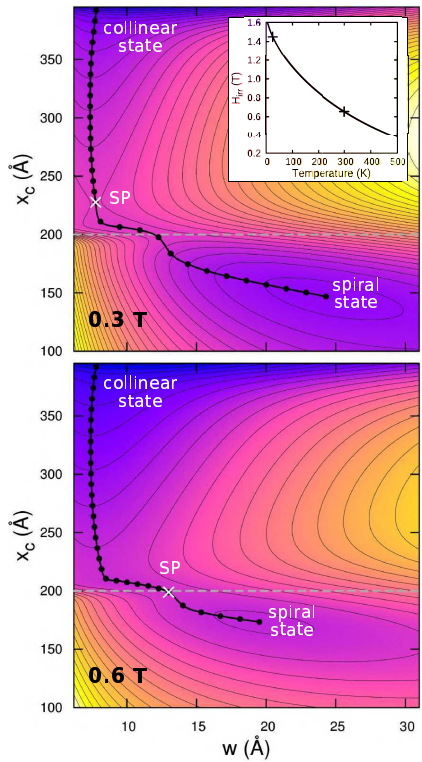}
\caption{
Energy surface for the Fe/Sm-Co spring magnet corresponding to applied magnetic field of H=0.3~T (upper) and H=0.6~T (lower),
constructed by assuming the magnetic moments of the atomic layers are oriented as described by Eqn.~(\ref{contplpaths}), 
mimicking a 
domain wall with width $w$ centered at $x_c$, and using the multilayer model of Eq.~(\ref{equ:model}).
The spiral state corresponds to the energy minimum to the right while 
the homogeneous, collinear state with parallel magnetic vectors corresponds 
to a minimum beyond the upper left corner.
Discretization points of the minimum energy path
are shown
by filled circles. The saddle points are identified with a X.
This graph illustrates how the magnetic field deforms the energy surface, destabilizing the spring state and
decreasing the activation energy for the transition, whereby the thermal excitation needed for the transition becomes smaller. 
The inset shows calculated values of H$_{irr}$ as a function of temperature, as well as two experimental data points from
the hysteresis loops in Fig.~\ref{fig1}.
}
\label{fig2}
\end{figure}

The energy of the system given by the multilayer model is a function of several hundred variables, 
one for each atomic layer.
In order to visualize this as an energy surface, we use a reduced description of the model in terms of only two essential variables.
This is accomplished by choosing a functional form for the orientation of the magnetic 
vectors
that mimics a 
domain wall (DW)
\begin{equation}
\label{contplpaths}
\phi_i = \phi(x_i) = \arctan{\left(\frac{x_i-x_c}{w}\right)} + \frac{\pi}{2} 
\end{equation}
were, $x_i$ is the location of layer $i$, $x_i = (i-1)d$, while $x_c$ is the location of the center and $w$ the width of the DW. 
This functional form is consistent with detailed calculations of the MEP using the full set of variables (see below).  
It can also describe the collinear state by placing the center well beyond the hard magnet.
With this model for the ordering of the magnetic moments, a contour graph of the energy surface can be constructed by inserting 
this expression for $\phi_i$ into the expression for the energy of the system, Eq.~(\ref{equ:model}). 
The resulting enegy surface obtained for two values of the magnetic field, H = 0.3 T and H = 0.6 T, is shown in Fig.~\ref{fig2}.
This visualization of the energy surface provides valuable insight into the mechanism of the transition.

An MEP for the transition between the two states is shown in Fig.~\ref{fig2}.
It reveals the mechanism of the transition and the displacement along the path defines a reaction coordinate. 
Starting from the local minimum corresponding to the spiral state, 
where the width of the DW is large, on the order of 25 layers,
the progression along the MEP involves a gradual decrease in the width of the DW as its 
center moves towards the interface between the hard and soft magnets.
After reaching the interface, the center of the DW remains there while the width decreases to ca.~5 atomic layers.
From then on, a narrow DW moves within the hard magnet away from the interface. 
The maximum increase in energy along an MEP gives the activation energy for the transition.
For a small field, this maximum (a saddle point on the energy surface) is within the hard magnet, 
but as the field is increased, it moves to the interface and stays there until the field reaches H$_{irr}$.
This can explain the experimentally observed sensitivity of the hysteresis width to the interface quality~\cite{Jiang05}.

As the field strength is increased, the local minimum corresponding to the spiral state becomes shallower and 
the activation energy for the transition from the spiral state to the collinear state decreases.
This is the primary reason for the narrowing of the hysteresis loop.  

\begin{figure}[h!]
\centering
\includegraphics[width=0.9\columnwidth]{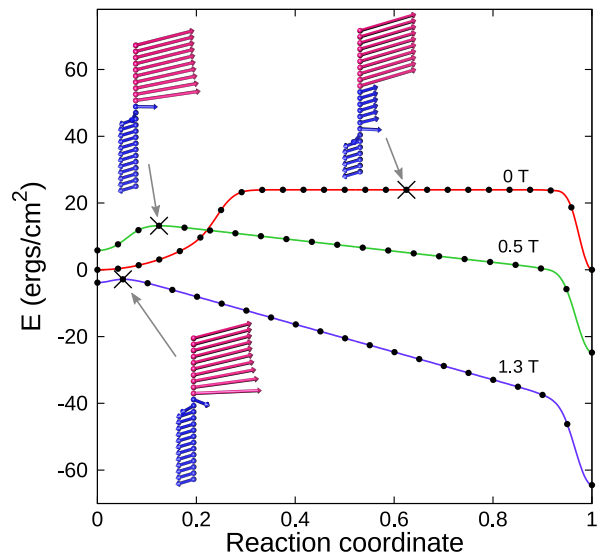}
\caption{
Calculated minimum energy paths for a transition from the spiral state in the soft magnet to the collinear state 
for two values of the applied magnetic field, H = 0.5 T and H = 1.3 T, as well as in the absence of a magnetic field.
The energy of the images in the geodesic nudged elastic band calculation for the model described by Eq.~(\ref{equ:model}) are shown 
with filled circles. The saddle points which give the activation energy are marked with X. 
The insets show the orientation of the magnetic moments at the saddle points.
}
\label{fig3}
\end{figure}

Results of calculations of the MEP for the full model (without the approximation in Eqn.~(\ref{contplpaths})) 
using the geodesic nudged elastic band (GNEB) method~\cite{Bessarab15}, 
an extension of the NEB method~\cite{NEB} to magnetic systems, are shown in Fig.~\ref{fig3}. 
When a small field is applied, the energy of the initial, spiral state increases because the magnetization 
has a component pointing in a direction opposite to the field.
For a stronger field, the energy of the spiral state decreases because more of the magnetic vectors in the soft magnet are aligned with the 
field leading to a magnetization component pointing in the direction of the field.   
The point of maximum energy along the MEP moves closer to the spiral state minimum
and becomes lower as the field increases, consistent with the contour graphs in Fig.~\ref{fig2}.
The height of the barrier is essentially the energy cost of moving the DW to the interface.
After the energy maximum has been reached, the 
energy changes linearly along the MEP as the center of the DW moves within the hard magnet. 
The width of the DW then remains unchanged (see Fig.~\ref{fig2}) so the exchange energy is constant and 
the  
slopes of the linear segments in Fig. 3 are 
determined by the Zeeman energy.
Since the magnetization of the system increases linearly with the reaction coordinate in this region, 
the energy changes linearly and the slope is larger for a stronger external field, 
a consequence of the stabilization of the collinear state by the applied field.

These results can now be used to estimate quantitatively how much the hysteresis loop narrows when
the temperature is increased.
The system will remain in the spiral state until the applied field has lowered the energy barrier sufficiently
and, thereby, decreased the lifetime of the spiral state sufficiently for the transition 
to the collinear state to occur on the time scale of the experiment. 
This approach is similar to the finite temperature micromagnetic method which has been used to calculate angular and thickness dependence of the coercive field for graded perpendicular recording media \cite{Suess11} .
The expression for the lifetime obtained from HTST is of Arrhenius form,
 (see Supplemental Material),
\begin{equation}
\tau = (1/\nu) \exp{(\EuScript{E}_a / {k_BT})},
\end{equation}
where the pre-exponential factor, $\nu $ turns out to be approximately temperature and field independent 
(see below) while the activation energy, 
$\EuScript{E}_a$ = $\EuScript{E}_a(H)$
obtained as the rise in energy along the minimum energy path, is strongly field dependent. 
Assuming constant pre-exponential factor, H$_{irr}$ can be determined by finding the field strength giving a certain
transition rate from the spiral state to the homogeneous state. 
From the Arrhenius expression, one obtains 
\begin{equation}
{{\EuScript{E}_a(H_{irr,1})} \over T_1 }= {{\EuScript{E}_a(H_{irr,2})} \over T_2} \ 
\end{equation}
an implicit expression showing how H$_{irr}$ changes with temperature.
Combined with calculations of minimum energy paths
(analogous to the ones shown in Fig.~\ref{fig3}), H$_{irr}$ is calculated to drop by a factor of 
0.45
as temperature is raised from 25 K to 300 K 
in excellent agreement with the experimental data shown in Fig.~\ref{fig1} (see inset).
This simple analysis based only on basic assumptions and the multilayer model for the energy of the system, Eq.~(\ref{equ:model}), 
can predict quantitatively the narrowing of the hysteresis loop with temperature 
without any temperature dependent adjustment of the model parameters.
A more detailed analysis including quantitative evaluation of the prefactor can, furthermore, give an estimate of
the value of H$_{irr}$ at a given temperature, as described below.

The pre-exponential, $\nu$, accounts for the entropy associated with vibrations of the magnetic moments. 
An HTST estimate of the pre-exponential factor in the Arrhenius rate law based on 
the
Landau-Lifshitz equations of motion 
can be obtained from the eigenvalues of the Hessian matrix at the initial state minimum
and at the saddle point~\cite{bessarab_12,Bessarab13a}.
The calculations for the present model of the Fe/Sm-Co spring magnet 
give a maximum of $\nu = 1.5 \times 10^{12}$ s$^{-1}$ at a low field 
of
 H= 0.2~T, and then a monotonically decreasing value as 
the field is increased down to $\nu = 2.6 \times 10^{11}$ s$^{-1}$ at a field of 1.6~T (see Supplemental Material). 

In order to estimate an absolute value of the rate, the relevant nucleation area needs to be estimated since the model in Eq.~(\ref{equ:model})
gives energy per unit area. Taking the layer thickness, $d$, to be 2 \AA, the relevant area within a layer where 
the saddle point configuration has been reached can be estimated from the observed H$_{irr}$ at 25 K.  
This analysis gives 2000 \AA$^2$ (see Supplemental Material), a value quite consistent with
atomic scale simulations of magnetization reversal in Fe islands where the 
width of the
temporary domain wall 
was 
found to be 30 \AA~\cite{bessarab_13}. The area obtained here is roughly this length squared.
This 
gives an activation enegy of $\EuScript{E}_a$ = 0.97~eV at 0.5~T and  0.13~eV at 1.3~T 
using saddle point energy values from 
MEP
calculations,
as the ones shown in Fig.~\ref{fig3}.
Combined with the estimate of the pre-exponential factor above, and assuming the time scale of the experiment is a second,
H$_{irr}$ comes out to be 1.45~T at 25~K and 0.65~T at 300~K.  
These values are in close agreement with the experimental measurments, as illustrated in the inset of Fig.~\ref{fig1}. Therefore, 
the change in the hysteresis loop with temperature can be described here
solely in terms of thermal activation in the system.

The analysis and calculated results described above show how important it is to consider thermally activated transitions 
in the analysis of the temperature dependence of magnetic hysteresis loops. 
A spring magnet has been taken here as an example because it is a well defined system but similar considerations apply more generally.

While there will be some change in
the average magnetic moment, $M$, with temperature, which will 
change the anisotropy energy~\cite{Callen66,Asselin10}, 
this effect is small unless the temperature is close to the Curie temperature. For both Fe and SmCo the decrease is 
on the order of 3\% as temperature is increased from 25 K to 300 K~\cite{Butler72,Gu10}.
In order to reproduce the ca. factor of 1/2 narrowing of the hysteresis loop 
observed experimentally (see Fig.~\ref{fig1}) 
by modifying only the anisotropy constant, 
$K$, for the hard magnet in Eq.~(\ref{equ:model}), a decrease to nearly 1/3 would be needed, 
well beyond the magnitude that a decrease in average magnetization would produce. 
The
large narrowing of the hysteresis loop is of fundamentally 
different origin, namely the increased rate of thermally activated transitions, as explained above.

\begin{acknowledgments}
This work was supported by the University of Iceland Doctoral Scholarship and Research Funds, 
the Russian Foundation of Basic Research (Grants No. 14-02-00102, and No. 14-22-01113 ofi-m), 
the Icelandic Research Fund, the Academy of Finland (grant 278260), 
and by the Nordic-Russian Training Network for Magnetic Nanotechnology (NCM-RU10121). 
PB gratefully acknowledges support from the G\"oran Gustafsson Foundation.
We thank Eric Fullerton for helpful discussions.
\end{acknowledgments}



\end{document}